\documentclass[12pt, lettersize]{article}
\usepackage{amsmath}
\usepackage{graphicx,psfrag,epsf}
\usepackage{enumerate}
\usepackage{natbib}
\usepackage{hyperref}

\newcommand{\blind}{1}

\begin{document}

\def\spacingset#1{\renewcommand{\baselinestretch}%
{#1}\small\normalsize} \spacingset{1}


\if1\blind
{\title{\large {\bf Probing the Stochastic Machine: Engaging with LLMs in Statistics Curricula Through Veridical Data Science}\\ Discussion of ``An Overview of Large Language Models for Statisticians'' by Ji {\em et al}.\ }
  \author{Tian Zheng\\
    Department of Statistics, Columbia University, New York, NY}
  \maketitle
} \fi

\if0\blind
{
  \bigskip
  \bigskip
  \bigskip
  \begin{center}
    {\large\bf Title}
\end{center}
  \medskip
} \fi

\subsection*{The Case for LLMs in the Statistics Curriculum}
\cite{ji2026overview} have produced the introduction to large language models (LLMs) that the statistics community needs. This review is notable for its breadth, spanning the training pipeline, alignment, uncertainty quantification, watermarking, fairness, interpretability, and self-alignment. More importantly, it consistently invites a statistical reading of one of the fastest moving phenomena in artificial intelligence beyond its engineering achievements. 

One observation that can be drawn from reading this comprehensive review is that the LLM is a system we have built whose behavior we do not yet fully understand. Its architecture, training data, and loss function can all be clearly articulated as in \cite{ji2026overview}. Given such a complete technical access to its mechanics, it is tempting to conclude that full understanding of LLMs should be attainable. Yet many of its most consequential behaviors, including in-context learning \citep{dong2024survey}, chain-of-thought reasoning \citep{wei2022chain}, calibration failures, and instability under prompt rephrasing, are observed empirically and explained only in part. The authors characterize the current state of the field as one of {\em partial empirical understanding}. Improving our understanding of LLM behavior is a necessary precondition for using these systems responsibly and beneficially. 

Part of what makes LLMs resistant to full characterization is that they are systems embedded in ongoing interaction and iterative updating. Formally, they are stochastic systems whose outputs during each generative step are samples from conditional probability distributions shaped by training, alignment, and inference, but whose full probabilistic structure remains only partially characterized. 
It is designed to be updated and re-aligned, and to evolve through iterative training on both human-generated data and model-produced outputs, as well as through ongoing interaction with users. As a result, LLM behavior emerges from a coupled system: human inputs shape model outputs, which in turn shape subsequent human perception and action. Natural language further complicates this interaction by acting as a lossy channel in both directions. It compresses user intent into prompts and model representations into generated text, which readers must reconstruct through their own interpretive frames. These transformations introduce systematic distortions that are often overlooked in practice. The review documents these effects primarily in Section 3.6 and Section 4.1, including stochasticity in decoding, biases in human raters and reward models, failure modes of LLM self-alignment evaluation, and gaps between model confidence and accuracy. 

Against this backdrop, Section 6.3 rightly frames the future of LLM use as one of human-AI collaboration and emphasizes intentional collaboration as the mechanism through which such systems can yield sustained improvement. This, in turn, highlights a {\em central gap}: how these interaction practices should be structured across users, from the general public to domain experts. This perspective motivates the present discussion. Stochastic systems whose behavior must be characterized empirically and that are continuously reshaped through human interaction are precisely the kinds of objects Statistics is equipped to study  \citep{NAP29292}. At the same time, LLMs are now the primary interface through which most people encounter AI. It follows that teaching students across levels of expertise to systematically probe and reason about LLM behavior is a natural foundation for AI education. \cite{ji2026overview}'s review outlines not only the technical landscape but also a framework for how such an education might proceed.

\subsection*{LLMs as Dynamical Systems: A Statistical Perspective for Curriculum Development}

To engage LLMs meaningfully in statistics education, we need a framework for structuring how students interact with and reason about their behaviors. A useful starting point is to treat the LLM, together with its users and training processes, as a dynamical system. Such systems may be fully specified at the level of local rules yet exhibit complex, difficult-to-characterize behavior when coupled to their environment through feedback. The LLM exhibits this strongly. The most visible feedback channel is human: prompts shape outputs, which in turn influence subsequent user behavior. As noted in Section 6.3, the data distribution in human-AI collaboration ``{\em adapts dynamically based on prior AI suggestions and human feedback},” introducing selection bias, concept drift, and strategic adaptation. This dynamic is closely related to performative prediction \citep{perdomo2020performative}, where model outputs feed back into and reshape the data-generating process. More broadly, such recursive interactions can shift output distributions over time, underscoring the dynamical nature of these systems and the statistical challenges they pose.

Framed this way, LLMs are both objects of statistical analysis and interactive systems that function as environments for statistical learning. Their stochastic behaviors can be accessed both through static inspections of model specifications and training algorithms, and through use: by prompting, perturbing, and comparing outputs across contexts. Through such interaction, statistical patterns become observable in the form of variability across runs, systematic differences across inputs, and feedback effects over time. This makes LLMs particularly well suited for teaching core statistical concepts.

This dynamical perspective also informs how human-AI collaboration can be structured in practice. LLMs now underpin a wide range of everyday technologies: chatbots and copilots, semantic search systems, agentic workflows that plan and execute tool use, and ``vibe coding,'' where natural-language descriptions are translated into working software. Designing learning activities that probe LLM behavior therefore offers a concrete way to cultivate statistical habits of mind that are critical in the age of AI \citep{NAP29292}. In particular, several challenges in contemporary AI use translate directly into opportunities for statistical inquiry.

First, \textit{fluency outruns accuracy}. LLMs are optimized to produce plausible text, not necessarily correct answers. As \cite{ji2026overview} emphasize, models can be confident and wrong simultaneously, and the model-generated uncertainty signals that might flag this are themselves unstable under perturbation. For the users, model confidence is not a reliable indicator of correctness. Pedagogically, this invites systematic probing: comparing outputs across variations in prompts, checking responses against known results, and treating answers as hypotheses rather than facts.

Second, \textit{bias emerges at the level of distributions}. As discussed in Section 4.5, prompts that differ only by gendered names can produce systematically different descriptions. Any single output appears unremarkable; the pattern becomes visible only across repeated samples. Bias, in this sense, is a statistical property, requiring reasoning about populations of outputs rather than individual responses. In the classroom, such effects can be surfaced through aggregation and comparison, turning interaction into data for analysis.

Third, \textit{interaction is inherently stochastic}. Outputs depend on sampling, training data composition (including contamination and annotation noise), reward-model design, and decoding parameters. As cataloged in Section 4.1, these sources of variation are substantial yet largely invisible to the user. Two responses to the same prompt are draws from a conditional distribution. Recognizing this shifts the mode of engagement from seeking a single ``correct'' answer to characterizing variability across runs.

These opportunities motivate the design of inquiry-based learning experiences in which students engage with LLMs through systematic experimentation. There has already been substantial discussion of how LLMs can be integrated into data science education, both as a tool to support students' learning and as part of a practicing data scientist's toolkit (e.g., \citep{tu2024should}). The present discussion focuses on an additional value: intentionally designing statistical investigations of LLM behaviors as part of formal statistics curricula. 

\subsection*{Veridical Data Science Offers a Design Framework}

The opportunities outlined in the previous section define a core set of competencies for engaging with LLMs. For non-experts, they form the basis of practical literacy: the ability to use LLMs critically and effectively. For statistics students, this extends further, motivating the development of expertise in designing human-AI workflows (e.g., via statistical ``wrappers'' for responsible LLM use; Section 6.1), systematically evaluating model behavior, and contributing to methodological and theoretical advances. This dual role aligns with recent calls from the National Academies. In Frontiers of Statistics in Science and Engineering: 2035 and Beyond \citep{NAP29292}, Conclusion 6-3 emphasizes the urgency of integrating statistical thinking into AI education for the public, while Conclusion 5-1 calls for sustained contributions from statisticians to evaluate and improve AI systems. \cite{ji2026overview} speaks to both audiences and invites a coherent, scalable framework for teaching across levels of the statistics curriculum. Under such a framework, engagement with LLM can be structured differently across audiences, but guided by a common set of principles that support both broad literacy and expert reasoning.

The Veridical Data Science (VDS) framework \citep{yuVeridicalDataScience2024} and its underlying Predictability-Computability-Stability (PCS) principles, first articulated in \cite{yukumbier2020}, provide such a framework. It offers a vehicle for embedding evaluation, reliability, and ethical reasoning into data science education \citep{NAP29292}. \cite{rosenberg2026teaching} provides a rich discussion of how to operationalize VDS/PCS in the classroom. In the following, I discuss how the PCS principles extend to issues specific to LLMs as discussed in the previous section.

For LLMs, \textit{predictability} is best understood at the level of distributions rather than individual outputs. An LLM generates text by sampling from a conditional distribution shaped by its training data. As a result, its outputs reflect patterns in that data, which may produce biased distributions that differ from what users intend. The relevant question is therefore not simply whether an answer is precise, but whether the distribution being generated aligns with the target of interest. 

In \cite{yuVeridicalDataScience2024}, \textit{computability} is largely a question of reproducible code. For LLMs, it is more fundamental: the chat interface obscures a set of computational constraints that shape model behavior in systematic ways. A \textit{computability} check, in this setting, asks whether the user understands these constraints and can trace their effects on outputs. Several are especially important. The \textit{context window} limits how much information the model can attend to, causing earlier content to be dropped in long interactions. Even within the context window, \textit{effective attention} could be uneven: models may underweight information in the middle of long inputs (“\textit{lost in the middle}”; \cite{salvatoreLostMiddleEmergent2025}) or disproportionately focus on certain tokens (attention sinks; \cite{guWhenAttentionSink2025}). In systems \textit{augmented with retrieval} or tools, pipeline boundaries introduce additional failure modes, where missing or incorrect retrieval leads the model to fall back on parametric memory without signaling it to the user. Finally, \textit{tokenization} shapes how inputs are represented, creating mismatches between user intent and model processing. Much of what appears as erratic behavior can be attributed to these computational constraints rather than to ``failures of reasoning.'' 

{\em Stability} asks whether conclusions are robust to reasonable perturbations, such as changes in prompts, models, or implementation details. For LLMs, instability is often visible: small changes in wording or configuration can produce materially different outputs. This variability is not merely a failure mode; it also signals where the model is uncertain. In this sense, stability serves both as an evaluation tool and as a mode of use. By systematically varying inputs and comparing outputs, users can probe the reliability of responses, identify sensitive dependencies, and distinguish stable patterns from artifacts of configuration. Engaging with LLMs in this way shifts interaction from accepting single outputs to analyzing distributions of behavior.

\subsection*{Curricular Examples}

Here I offer four examples in the following to illustrate what VDS-informed teaching about LLMs can look like at different points on the educational ladder. They span teaching statistical literacy through research training in Statistics.

\paragraph{Example 1 (statistical literacy): ``ask it twice."}
For learners developing statistical literacy, the goal is to build intuition about LLMs through a statistical lens. A simple activity illustrates this. Students ask a chatbot, ``Tell me a short joke about dogs,'' record the response, then repeat the same prompt in a fresh interaction and compare the outputs. Why do they differ? The contrast introduces a key idea: LLM outputs are samples, not fixed answers. Extending the prompt, for example, by modifying the word ``\textit{joke}'' to ``\textit{funny story}'' further reveals the conditional nature of the system.

Viewed through a dynamical lens, each interaction can be treated as a draw from a system shaped by stochasticity and context: restarting or continuing a session produces different trajectories. Students can be guided to form hypotheses and design simple experiments to test them. Using terms from the Guidelines for Assessment and Instruction in Statistics Education (GAISE) Reports \citep{franklin2020introducing}, this is variability encountered through interaction; in VDS terms, it is a stability check, probing how outputs change under repetition or small perturbations.

\paragraph{Example 2 (Introductory Statistics): asking an LLM for random numbers.}

In an introductory Statistics course, consider a simple exercise in which students repeatedly ask an LLM to ``produce a sequence of 30 random numbers drawn from the digits 0-9,'' collect the responses, and tabulate or visualize the resulting frequencies. This activity highlights how the term ``random'' is interpreted by the model as part of the conditioning for generation. Through experimentation, students can observe that the outputs are not uniform or independent; Some digits (e.g., 7) are overrepresented, while others are suppressed. Long sequences of digits avoid repeats. These interesting patterns provide a concrete illustration of \textit{distributional predictability}: the model is generating from a learned conditional distribution that reflects patterns and ``assumptions'' in its training data rather than the intended uniform distribution. The exercise also offers a setting in which students can study LLM behaviors using formal statistical tools.

\paragraph{Example 3 (professional education, e.g., law): Is the LLM a good super-judge of ordinary meaning?} For professional schools such as law, medicine, business, and public policy, discussions on AI education focus on developing a working understanding of how LLMs can produce fluent, authoritative outputs while computing a quantity different from the one users care about. Legal scholarship provides a clear illustration.\footnote{This example was brought to the author's attention by Professor Rebecca E.\ Wexler at the Columbia Law School, who is an expert at the intersection of law and technology. She has been collaborating with the author as an aiX Faculty Fellow, \href{https://provost.columbia.edu/news/faculty-led-initiative-aix-faculty-fellowship-program}{a faculty-led initiative on AI education at Columbia}.} Recent work on generative interpretation proposes using LLMs to resolve interpretive questions \citep{hoffman2024generative}, while \cite{grimmelmann2025generative} identifies two central issues: a reliability gap, where outputs vary across models and prompts, and an epistemic gap, where the model estimates a conditional probability over training text rather than over legal meaning.

These critiques map naturally onto the Predictability-Computability-Stability (PCS) framework \citep{yukumbier2020}. The reliability gap reflects a lack of stability: outputs are sensitive to perturbations in prompts and model choice. The epistemic gap concerns predictability: the model may produce consistent estimates, but for a target that is misaligned with the quantity of interest. A single class session can make this concrete by comparing outputs across prompts or model versions and asking what quantity the model is actually computing versus what users assume it computes. For statistics students, the same example can be extended into a mini-project. Students can assess stability under perturbations and clarify the predictability target being estimated, turning a domain-specific debate into a setting for studying LLM training, evaluation, and alignment.

\paragraph{Example 4 (statistics students): A PCS stability audit of LLM-based data-analysis tools}

For graduate statistics students, the goal shifts from using LLMs to interrogating them and identifying open research directions. Recent work on PCS workflows for veridical data science emphasizes studying variability in workflow decisions made throughout the data science life cycle (DSLC) \citep{rewolinski2025pcs}. Students can treat LLM-based tools as components within a DSLC, probe their behavior under controlled perturbations, and document how design choices propagate to downstream outcomes. A natural first exercise is a small-scale VDS-style stability audit. Students can assign a multi-step analytical task (data loading, preprocessing, modeling, and reporting) to one or more LLM tools, repeat the workflow multiple times, and introduce controlled perturbations (e.g., prompt variation or data splits). The resulting differences across runs reveal how model behavior depends on system design and interaction structure. 

%

Auditing LLMs under different computational settings can expose the engineering layer of LLM systems. Students may observe that smaller or resource-constrained models produce more erratic outputs, and explore how techniques such as retrieval-augmented generation (RAG) or modular agent design can mitigate this variability. In this way, the activity connects statistical reasoning with system design, giving students hands-on experience with how training, alignment, and architecture shape model behavior. The result is not only a deeper understanding of LLM methodology but also a foundation for developing efficient, reliable AI workflows under realistic computational constraints.

\subsection*{An Opportunity for Statistics in the Age of AI}
LLMs are widely used in our society, but the shared understanding of their behavior remains limited. Statistics is well positioned to shape that understanding, as many open questions about LLM behavior, such as uncertainty, robustness, and feedback, are inherently statistical. Meeting the emerging needs of AI education is therefore both a responsibility and an opportunity for the field \citep{NAP29292}. Teaching these systems can also surface phenomena that motivate further research.

\cite{ji2026overview} can be read as a blueprint for what students should learn to anticipate and probe in LLM behavior. This discussion has framed LLMs as stochastic systems whose behavior creates natural opportunities for teaching statistical reasoning. The VDS framework provides a principled way to organize that reasoning across levels of the curriculum, and the examples above are offered to illustrate how this integration can be realized in practice.

\bibliographystyle{apalike}
\bibliography{Bibliography-MM-MC}
\end{document}